\documentclass[useAMS,usenatbib]{mn2e}

\usepackage{graphicx}

\title[The Virgo over-density]{Stellar over-densities in the halo: the extent of the Virgo over-density\thanks{Based, in part, on observations obtained from Mount Stromlo's 1.3 meter Telescope (Great Melbourne Telescope).}}
\author[Keller et al.]{Stefan C.~Keller$^{1}$\thanks{E-mail:
stefan@mso.anu.edu.au}, Gary S.~Da~Costa$^{1}$ and Sayuri L.~Prior$^{1}$\\
$^{1}$Research School of Astronomy and Astrophysics, Australian National University, Cotter Rd, Weston, ACT 2611, Australia}
\begin{document}

\date{}

\pagerange{\pageref{firstpage}--\pageref{lastpage}} \pubyear{2008}

\maketitle

\label{firstpage}

\begin{abstract}
We map the three dimensional extent of the Virgo Over-density by combining distance information from RR Lyrae variables and projected spatial information from SEKBO \citep{Keller08} and Sloan Digital Sky Survey (SDSS) DR6 photometry. The Virgo Over-density is seen to comprise two filaments $14.5^\circ \times 3^\circ$ and $10^\circ \times 3^\circ$ and a circular structure $3^\circ$ in diameter.  Together the three features span 38$^{\circ}$ of right ascension and declinations of +2 to -15$^{\circ}$. RR Lyrae variables place the two filamentary features at heliocentric distances of 20 and 17 kpc respectively, with projected dimensions of $5\times1$ kpc and $3\times1$ kpc.
\end{abstract}


\section{Introduction}

One of the most important processes yet to be understood by modern astrophysics is that of galaxy formation. The halo of the Galaxy offers arguably the best opportunity to study galaxy formation due to its relative quiessence and our ability to obtain accurate kinematics and detailed chemical abundances for a large number of stars. The seminal work of \citet{ELS} proposed galaxy formation occurred in the wake of a monolithic collapse of a proto-galactic cloud. Later, \citet{SZ} challanged this notion by proposing that the halo formed via the accretion of numerous small entities over an extended period.

Today, galaxy formation is discussed within the context of LambdaCDM cosmology. This predicts hierarchical structure formation in a way that qualitatively parallels the Searle \& Zinn scenario. Observational evidence of the duality in the Galactic halo \citep{Freeman02, Carollo07, Bell07} suggests at least two formation mechanisms. The kinematics and star formation history of the inner halo ($R<$20 kpc) suggest formation from rapid collapse at an early epoch. The outer halo, on the other hand, possess kinematics \citep{Carollo07} and a level of substructure \citep{Bell07} consistent with assembly from disrupted dwarf galaxies.

\subsection{The Sagittarius dwarf and the Virgo over-density}
The most prominent on-going accretion event in the Milky Way halo is the dissolution of the Sagittarius dwarf \citep[Sgr, ][]{Ibata94}. The leading and trailing arms of Sgr tidal debris are seen to wrap around the sky \citep{Majewski03, Newberg02, Newberg07, fieldofstreams, Keller08}. The debris of Sgr are the largest contribution to substructure in the outer halo yet to be found.

Arguably the next most significant outer halo substructure is the Virgo Over-density (VOD). The VOD was discovered in the Sloan Digitial Sky Survey \citep{York00} as a diffuse over-density of F-type main-sequence stars spanning some 1000 deg$^{2}$ \citep{Newberg02,Newberg07} and distances of 6-20 kpc \citep{Juric08}. The region has a coincident series of over-densities in RR Lyrae variable stars \citep{VivasZinn06, Keller08} at distances of 16-19 kpc. The studies of \citet{Duffau06} and \citet{Prior08} found subsets of RR Lyraes show a common radial velocity with very low velocity dispersion (consistent to measurement uncertainties). This is as expected from kinematically cold tidal debris. The moving group was termed the Virgo Stellar Stream to distinguish it from the general stellar over-density.

The nature of the VOD remains uncertain. Does it represent a separate accretion event, or is it related to the spatially overlapping Sgr debris? \citet{MartinezDelgado07} proposed that the VOD results from the confluence of the leading and trailing arms of Sgr as seen in the n-body models of \citet{Law05}. The Law et al.\ models predict highly negative radial velocities for the Sgr material in this region. This is contrary to the observations of \citet{Duffau06} and \citet{Newberg07} who find most stars lie at a radial velocity of $+$100-130 kms$^{-1}$. \citet{Newberg07} also notes that the models fail to predict the high stellar density seen in the VOD, and that the VOD should be spatially offset from the Sgr leading arm as we observe. This leads \citet{Newberg07} to conclude that the VOD is a halo substructure of separate origin to Sgr. \citet{Vivas08} examines the radial velocities of a sample of RRLs in the direction of the VOD at distances of less than 13 kpc. \citeauthor{Vivas08} finds that the Virgo Stellar Stream extends to $\sim12$ kpc, but none of the additional RRLs can be ascribed to the leading arm of Sgr. Nevertheless, the study of \citet{Prior08} find RRLs at negative radial velocities which they speculate could be Sgr debris \citep{Prior08b}.

It is evident from \citet{Newberg02, Duffau06, fieldofstreams, Newberg07, Juric08} that the centre of the VOD lies somewhere to the south of the region surveyed by SDSS. The VOD was traced southwards by \citet{Keller08}. They identified two RR Lyrae over-densities in the VOD region: one at a heliocentric distance of 16 kpc, 8$^{\circ}$ south-east of the centre of the VOD defined by \citet{Duffau06}, and another at 19 kpc and 24$^{\circ}$ south-east of this centre. \citet{Prior08} explore the extent of the VOD by looking at a sample of SEKBO fields that sparsely sample the vicinity of the VOD. This study reveals that the VOD is a large diffuse feature, covering at least 760 deg$^2$ of sky.

In the present study, we draw together the SEKBO and SDSS data sets to constrain the 3-d extent of the VOD with the aim of clarifying its nature. We use the distance information available from the SEKBO RR Lyrae candidates and the spatial extent determined from the combined SEKBO and SDSS DR6 data sets. In section \ref{section:conc} we conclude with some remarks regarding the nature of the VOD.

\begin{figure*}
\begin{center}
\includegraphics[scale=0.7, angle=0]{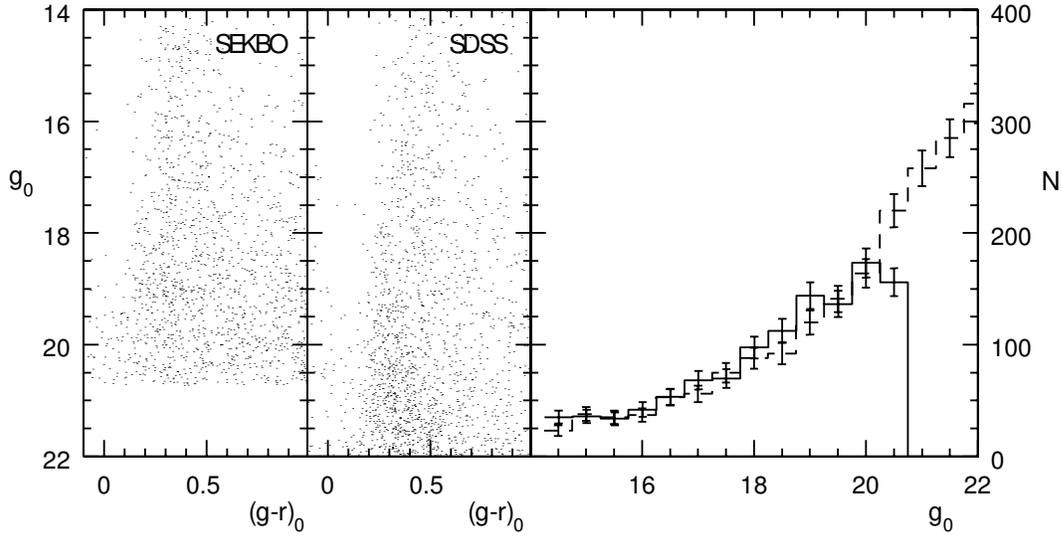}
\caption{{\bf{Left}}: A comparison of SEKBO and SDSS colour-magnitude diagrams for a 1$^{\circ} \times 1^{\circ}$ region centred on (RA, Dec) 192$^{\circ}$,-4$^{\circ}$. {\bf{Right}}: a comparison of the luminosity functions derived from the SEKBO (solid line) and SDSS (dashed line) fields with Poisson uncertainties shown. Note the agreement between the luminosity functions derived from the two data sets is within uncertainties until the completeness limit of the SEKBO data at $g_0 \sim 19.75$. }\label{figure:CMD}
\end{center}
\end{figure*}

\subsection{Luminosity function excess}
We utilise the Sloan Digital Sky Survey Data Release 6 \citep{Adelman-McCarthy08} and SEKBO \citep{Keller08} photometric databases to construct luminosity functions over the combined survey area. We have transformed the Johnson $V$,$R$ of the SEKBO survey \citep{Keller02} to SDSS $g$,$r$ for consistency using the transformations of \citet{Lupton05}. In order to uniformly combine luminosity functions from  the two datasets we needed to consider magnitudes that do not suffer from substantial observational incompleteness. As can be seen from the example colour-magnitude diagrams in Figure \ref{figure:CMD}, observational incompleteness becomes appreciable in the SEKBO data set at $g_0 \sim 20$ (there is some variation in depth amongst the SEKBO fields). This is almost 2 magnitudes brighter than the limit of SDSS photometry.

To construct the luminosity functions seen in Figure \ref{figure:CMD}, we utilise the colour range $0.1<(g-r)_{0}<1.0$ and magnitude range $14<g_{0}<19.75$. Reddening is as estimated from the \citet{Schlegel98} reddening maps. This colour range excludes the near-by red disk dwarf star sequence that is not of interest to the present study. The size of individual SEKBO fields ranged from 0.81 to 0.60deg$^2$. The areal coverage of each field was dependant on the dither pattern and operational status of CCDs. The SDSS data are compared over a 1deg$^2$ field. To correct for the variation in sample size, we normalised the luminosity functions over the magnitude range $14.5<g_0<15.5$. As can be seen from an example overlap region in Figure \ref{figure:CMD}, the SDSS and corresponding SEKBO luminosity functions agree within uncertainties.

We then compared the observational data with synthetic luminosity functions derived from the Besancon galaxy model \citep{Robin03}. The simulations were constructed for 1deg$^2$ fields with a distance range of 0-120 kpc with magnitude and colour cuts applied as per observations. The Besancon model is a dynamically self-consistent, parametric description of the Galaxy. As \citet{Juric08} note, a parametric model is susceptible to the choice of input observations since these are typically based upon small areal coverage and many include unrecognised stellar substructure. The solution of \citet{fieldofstreams} and \citet{Juric08} is to directly map the stellar number density using distances derived for main-sequence stars from photometric parallax. The technique requires highly accurate colours in order to precisely derive an absolute magnitude. \citet{Juric08} show, magnitudes determined to a precision of 0.01-0.02 mag.\ result in an uncertainty in distance of $\pm10$\%. This is a result of the extremely steep absolute magnitude - colour relation inherent in the photometric parallax method. In addition, considerable systematics enter with the choice of a photometric parallax relation (up to $\pm70$\% in distance for the bluest halo stars).

Our approach to discerning halo substructure is unlike that utilised by \citet{fieldofstreams} and \citet{Juric08} that have focused on the density of f-type main-sequence stars. The SEKBO data lacks the  global photometric accuracy required for such a technique. Due to propagation of uncertainties in standardisation and colour transformation from the non-standard MACHO B \& R passbands, the global accuracy of SEKBO phtometry is $\pm0.03$ magnitudes in $V$ \& $R$ bands \citep{Keller08} for $g<16$. This rises to $\pm0.15$ magnitudes at $g\sim19.75$ magnitudes ( with some variation from field to field due to variable survey depth).

For this reason, we have chose to construct our search for halo substructure via the luminosity function of the obsevred data. This approach has also been applied in the study of \citet{Duffau06}. \citeauthor{Duffau06} has examined the region of sky between 176$^\circ$$\le$RA$\le$210$^\circ$ and -4Ã$\le$Dec$\le$+4$^\circ$ in 2$^\circ$$\times$2$^\circ$ fields. The outline of significant over-density as described by \citet{Duffau06} is shown in Figure \ref{figure:NGC}. Our approach has also been applied by \citet{Newberg07} to two fields, one centred on RA=186$^{\circ}$ and Dec=0$^{\circ}$ and one further south at RA=191$^{\circ}$, Dec=-7.8$^{\circ}$.  Both fields reveal significant luminosity function excess relative to fields at the same galactic latitude but 180$^{\circ}$ displaced in longitude. Nonetheless, the disadvantage of the our technique has the drawback of removing $distance$ information from the view of substructure. However, in Section \ref{section:RRL} we merge distance determinations from the distribution of RR Lyraes to constrain the distance to the substructures.

\begin{figure*}
\begin{center}
\includegraphics[scale=0.7, angle=0]{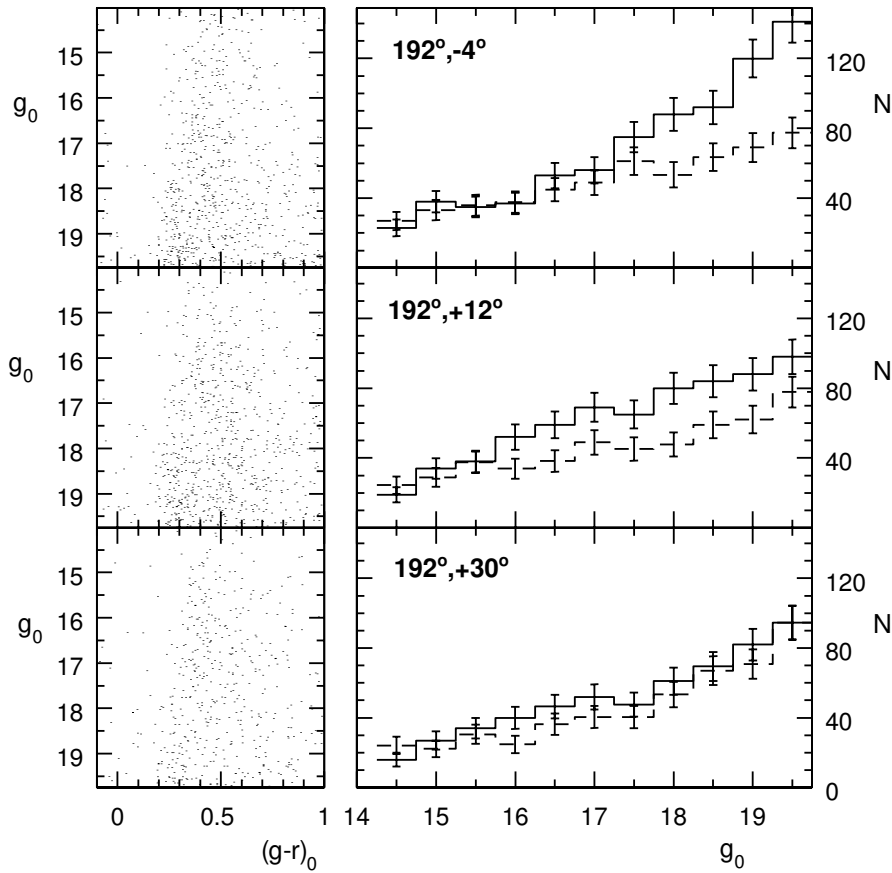}
\caption{Three 1$^{\circ} \times 1^{\circ}$ fields progressively further away from regions of marked excess. The top row shows a field centred on the VOD (RA, Dec) 192$^\circ$, -4$^\circ$ and the middle and bottom rows are for fields consecutively 8$^\circ$ north \citep{Keller02}. The right panels compare the luminosity function observed (solid line) with that predicted from the Besancon Galaxy model \citep[dashed line,][]{Robin03}.}\label{figure:excess}
\end{center}
\end{figure*}

Localised excess over the Besancon model shows regions that depart from an axially symmetric model. While features covering several thousand square degrees (quad\-rupole moments, for instance) may suffer from uncertainty due to the validity or otherwise of the model, we can consider features on scales of several hundred square degrees as representing departures from the local smooth background density. 

The bottom panel of Figure \ref{figure:excess} shows a field well separated from substructure in the southern Galactic cap. Here we see the agreement of the observed LF with the Besancon model LF to within the Poisson uncertainties. As we proceed through to the area of the VOD there is a marked excess of stars rising towards the faintest bin considered (at $g_0$=19.5 in Figure \ref{figure:excess} top, see also \citet{Prior08}). 

The cumulative difference between the observed and predicted luminosity functions is then expressed in terms of significance relative to the Poisson uncertainties inherent in the observed sample. That is,
\begin {equation}
S = \Bigl( \sum_{i} \bigl[ (N_{obs} -N_{model})_{i} /  \sqrt{ ( {N_{obs}^{2}} + {N_{model}^{2}} ) }_{i} \bigr]^2 \Bigr)^{1/2}
\end{equation}
where $S$ is the significance of a field, $N_{obs}$ is the observed number of objects to magnitude bin $i$, $N_{model}$ is the predicted number of objects to bin $i$ from Besancon galaxy model normalised to that observed in the magnitude range 14.5$ < g_0 < 15.5$.

\begin{figure*}
\begin{center}
\includegraphics[scale=0.55, angle=0]{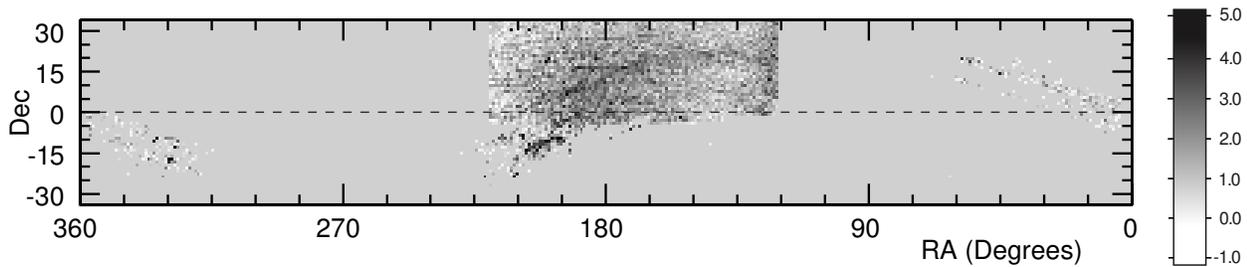}
\caption{The significance of cumulated excess of the observed luminosity function compared to that predicted by the Besancon galaxy model \citep{Robin03} to a limiting magnitude of $g_0$=19.75 in 1$^\circ \times 1^\circ$ fields. Darker shading represents more significant excess over the model, as described by the greyscale colour bar to the right. The significance is expressed as the number of sigma over the model (see text for details). }\label{figure:skysignif}
\end{center}
\end{figure*}

\begin{figure*}
\begin{center}
\includegraphics[scale=0.22, angle=0]{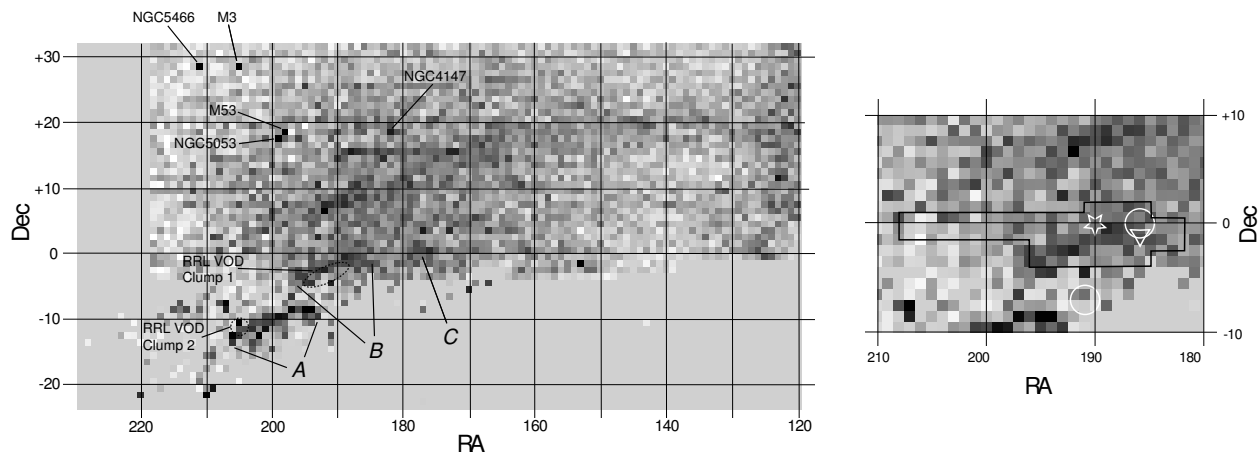}
\caption{{\bf{Left:}} A close-up of Figure \ref{figure:skysignif} to show the central ``Field of Streams'' region. Numerous point-like features are associated with known globular clusters as shown. The leading arm of Sagittarius debris is seen as the broad feature from 120$^{\circ}$,+20$^{\circ}$ to 190$^{\circ}$,+8$^{\circ}$. Three new features $A$, $B$, and $C$ are indicated. The dashed ellipses mark the regions of excess RR Lyrae density from \citet{Keller08}.
{\bf{Right:}} A further enlargement of the region around the Virgo Over-density (Feature $B$) to show previous detections by \citet{Duffau06} (solid outline), \citet{VivasZinn06} (triangle), \citet{Ivezic05} (star) and \citet{Newberg07} (circles).}\label{figure:NGC}
\end{center}
\end{figure*}

The resulting map of significance of departure from the luminosity function predicted by the Besancon model is seen in Figure \ref{figure:skysignif}. The SDSS data cover the region of the `Field of Streams' of \citet{fieldofstreams}, namely 120$^{\circ}<$RA$<220^{\circ}$ and from the southern SDSS limit (varies from Dec=-1$^{\circ}$ to -4$^{\circ}$) to Dec=+34$^{\circ}$. In the region of overlap between SDSS and SEKBO we utilise the SDSS dataset in light of the superior photometric precision of the SDSS data.

Figure \ref{figure:NGC} expands the northern region of overlap between SDSS and SEKBO. A series of `point' sources are seen - these mark globular clusters in the field (distances $<100$ kpc, after this the horizontal branch falls out of the magnitude range considered for the luminosity function). The main source of the over-density in the north is the leading arm of Sgr \citep[see for example, ][]{fieldofstreams, Fellhauer06}. Enmeshed with the linear feature of the Sgr leading arm is a large, diffuse region of excess between RA=170-195$^{\circ}$ and Dec$<$+7$^{\circ}$. This diffuse region is also coincident with the broad VOD detection of \citet{Juric08} that is seen to cover $\sim$1000 deg$^{2}$(centred at $192^{\circ}$, $+2^{\circ}$). 

Three significant new features are also seen in the extension to the south-east afforded by the SEKBO data. A linear feature $A$ (198$^{\circ}$,-10$^{\circ}$) forms a band approximately $3^{\circ}$ wide and $14.5^{\circ}$ long. A second linear feature $B$ (192$^{\circ}$,-2$^{\circ}$), is $3^{\circ}$ wide and $10^{\circ}$ long. The third significant feature $C$, seen at (178$^{\circ}$,0$^{\circ}$) is approximately 3$^\circ$ in diameter. The three features are broadly consistent with the area of over-density discussed in \citet{Prior08}. 

\section{Distances to substructures from RR Lyrae variables}
\label{section:RRL}
The above detections of halo substructure are based upon the excess or deficit of main-sequence and red giant branch populations over the Besancon galaxy model. The range in luminosity exhibited by both stellar populations does not afford any distance information at the low surface densities seen here.

The \citet{Keller08} study of the RR Lyrae over-density from SEKBO presents the radial distribution of RR Lyrae candidates \citep[76$\pm7$\% of which are RR Lyraes,][]{Prior08} across the VOD region. Using the \citet{Keller08} data set, we can construct a plot of the number density of RR Lyrae candidates as a function of distance for the range in right ascension 170-210$^{\circ}$. 

The calculation of the RRL number density follows the technique developed by \citet{Wetterer96} and implemented in \citet{Keller08} (to which we refer the reader for details). The following equation describes the local space density as a function of galactocentric distance:
\begin{equation}\label{equ:fofr}
\rho(R) = \frac{1}{4 \pi R^{2} f(R)} \frac{dN}{dR},
\end{equation}
where $f(R)$ is the fraction of the total halo volume at $R$ that is sampled by the survey ($f(R)$ is analogous to a solid angle) and $N$ is the number of RRL as a function of distance. Whereas the solid angle is constant as one looks through the halo, $f(R)$ varies as a function of Galactocentric radius and hence must be calculated numerically for each field. To account for the effects of completeness we multiply each $f(R)$ by the Monte-Carlo derived RRL$ab$ completeness profile for the field as a function of galactocentric radius. This gives the \emph{effective} volume of the halo that is sampled by each field. We use Equation \ref{equ:fofr} to calculate the local space density for each candidate. A transformation to heliocentric distance is then made.

\begin{figure}
\begin{center}
\includegraphics[scale=0.37, angle=0]{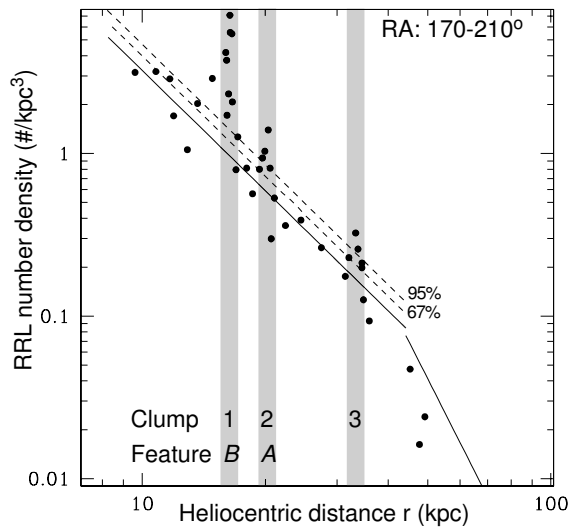}
\caption{A plot of the number density of RR Lyrae candidates in the RA range 170-210$^{\circ}$ as a function of heliocentric distance. RR Lyrae candidates are from \citeauthor{Keller01} (2008). The solid line shows the average power law relation of halo density. From Monte-Carlo simulations we are able to derive 1 and 2 $\sigma$ intervals (dashed lines) above this average relation based on random samples of appropriate sample size. The grey shaded regions indicate the locations of three significant over-dense regions.}\label{figure:clump}
\end{center}
\end{figure}

The result is seen in  Figure \ref{figure:clump}. From Monte-Carlo simulations of appropriate sample size using the halo density profile defined in \citet{Keller08} (the solid line in Fig.\ \ref{figure:clump} for $R<$45 kpc) we have determined the 1 and 2$\sigma$ significance contours. These are shown in Figure \ref{figure:clump} as the dashed lines. Figure \ref{figure:clump} shows three peaks of greater than 2$\sigma$ significance; at 17, 20 and 34 kpc (heliocentric).

From \citet{Keller08}, figure 20, we see that the VOD Clump 1 spatially extends over the linear feature $B$ and Clump 2 is seen to coincide with the southern portion of feature $A$, as shown in Figure \ref{figure:NGC}. The more distant, and less numerous, Clump 3 does not present a well defined spatial centroid. Feature $C$ does not exhibit a corresponding excess in RRLs. At their determined distances feature $A$ (20 kpc) has projected dimensions of $5\times1$ kpc and feature $B$ (17 kpc) dimensions of $3\times1$ kpc.

\section{Discussion of substructures}

An extensive literature has developed that examines the stellar over-densities in this region. Our study is the first to resolve the VOD region into a series of distinct substructures.  The filling factor of the three features is significantly less than the 760 deg$^2$ determined by \citet{Prior08} or $\sim1000$deg$^2$ by \citet{Juric08}. We believe this can be understood by the spatial sampling of \citeauthor{Prior08} and their extrapolation to an elliptical morphology. The three features span some 38$^\circ$ of sky similar to the major axis size of 45$^\circ$ determined by \citeauthor{Prior08}. It is by increasing our spatial resolution that we are now able to see the details of substructure in this region.

In the following section we discuss our finds in the light of the existing literature for the region.

\subsection{Feature $A$}

\citet{Newberg07} examines a 1.5$^{\circ}$ radius field centred at RA=191$^{\circ}$, Dec=-7.8$^{\circ}$. this corresponds to the peak overdensity seen in the southerly `outrigger' scan of SEGUE. This field overlaps with the north-west corner of feature $A$. SDSS spectroscopy reveals an marginally significant excess at radial velocities between V$_{GSR}$=+100-+130kms$^{-1}$. Such an excess is also seen by \citet{Newberg07} towards the VOD (feature $B$, see below). This suggests that feature $A$ and the VOD share a common spatial velocity and potentially a common origin.

\subsection{Feature $B$ - the VOD}

Feature $B$ is identified as the VOD. The VOD corresponds to the ``12.4hr'' clump described by \citet{VivasZinn06} and first detected as an overdensity in RRLs by \citet{Vivas01}. It was independently seen as an overdensity of F-type main-sequence stars by \citet{Newberg02} at RA=190$^{\circ}$, Dec=0 and heliocentric distance of 18 kpc. Using a 2.3$^{\circ}$ wide band centred on declination -1:10:48, \citet{VivasZinn06} find a peak density located at RA=186$^{\circ}$ and a heliocentric distance of 17 kpc.

Feature $B$ is also recovered in the study of RRL candidates from SDSS by \citet{Ivezic05}. The study of \citet{Ivezic05} finds the VOD centred at RA$\sim$190$^{\circ}$ in a 2.5$^{\circ}$-wide band centred on Dec=0$^{\circ}$. \citet{Ivezic05} derive a mean heliocentric distance of 18 kpc from the luminosity of the clump candidate RRLs.

\citet{Duffau06} presents spectroscopic follow-up of the ``12.4hr'' clump and finds 9 of 18 RRLs studied share a common radial velocity of +100kms$^{-1}$ with velocity dispersion less than measurement uncertainties. This moving group was termed the Virgo Stellar Stream to differentiate the population from the region of over-density. In an examination of SDSS photometry for the adjoining region of sky, \citet{Duffau06} saw evidence for an excess in the luminosity function (the same technique as applied in the present study) over an area of at least 106 square degrees of sky centred on RA=186$^{\circ}$ and Dec=-1$^{\circ}$. 

Since the over-density is located at the southern edge of the SDSS field the spatial extent remains unbound. In \citet{Newberg07} one of the three SEGUE `outrigger' scans apparently shows that the VOD extends beyond the Dec=-4$^{\circ}$ limit of SDSS to Dec$<-15$$^{\circ}$. In addition, \citet{Newberg07} examines the radial velocities of stars in a 1.5$^{\circ}$ radius centred in RA=186$^{\circ}$, Dec=0$^{\circ}$. At magnitudes corresponding to main-sequence members of the VOD with a distance of 18 kpc there is a suggestion of a population of stars with radial velocities between +100-+130 kms${-1}$ as seen in the study of \citet{Duffau06}. \citet{Vivas08} presents evidence that the VSS population extends towards us to around 12 kpc.

The present study places the VOD in a clearer spatial context by providing broader sky coverage. Our derived mean distance of 17 kpc is in good agreement with previous studies. As can be seen in Figure \ref{figure:NGC}, there is also agreement between previous positions of VOD presented in the literature and our map of over-density in this region.  

The discovery by \citet{Walsh08} of VirZ, a dwarf galaxy candidate at the north-western tip of feature $B$ is intriguing. Walsh et al.\ find an optimal distance of $\sim40$ kpc but considerable uncertainty is accommodated by the small number of stars populating the giant branch. The possibility of a connection between VirZ and feature $B$ is one that must await verification from deeper photometric and kinematic studies.

\section{A field of complexity --- the nature of the Virgo over-density}
\label{section:conc}
There are a number of possible interpretations for the morphology we see in Figure \ref{figure:NGC}. Structures $A$, $B$ \& $C$ could represent individual accretion events or they may be `hotspots' of increased density along a single stream. If describing a coherent structure, the distances to features $A$ \& $B$ do not indicate a significant line of sight inclination for the stream. 

The proximity of the features to Sgr and their alignment parallel to the Sgr leading arm stream could point to an origin from Sgr. In this scenario, $A$, $B$, \& $C$ may trace the trailing arm of Sgr some $\sim270^{\circ}$ downstream from Sgr.  All n-body models in the literature \citep{Law05, Helmi01, Fellhauer06}, while they differ on the distance to which the trailing arm extends, place the returning trailing arm at a distance of $\sim20$ kpc over the RA range of features $A$ \& $B$. However, these models do not predict the substantial stellar density seen in the two features. There is also no evidence for a return of the trailing arm from the distribution of M-giants. However, this material is expected to be derived from disruption four orbital periods, or (3-3.5) Gyrs ago \citep{Law05}, and given that the the bulk of M-giants are 2-3 Gyr or younger \citep{Layden00, Keller99} we would expect a marked reduction in the contrast of M-giant tracers along this section of stream.

The possibility of a common origin for features $A$, $B$, \& $C$ in Sgr is one that requires clarification as this is critical for studies of the orbit of Sgr and, flowing from this, our ability to constrain the shape of the dark matter halo. Once we have clarified what material is associated with Sgr, we can then realise the potential of Sgr as a test particle of the Galaxy's dark matter halo shape, $q$($\equiv c$/$a$). $q$ is fundamental to Galactic dynamics. Recently, the RAVE consortium \citep{Siebert08} was unable to estimate $q$ due to uncertainties in halo+disk mass and disk scale length. Considering the distance and extension of the Sgr debris, they represent the best opportunity to constrain $q$. Tracing Sgr debris through the VOD region will provide substantial new kinematic information that will allow much tighter constraints on the models. In turn, this may resolve the discrepancy in determination of $q$ (current literature ranges from oblate $q=0.8$, to prolate $q=5/3$).

Should subsequent observations and modelling reveal that the substructure revealed by the present study originates from Sgr, this will represent an important constraint on the number and mass of accretion events in the outer halo over the last 3-4 Gyrs. The accretion history may then be compared to expectations from cosmological simulations. The small scale height of the disk suggests an anomalously low late-time merger rate \citep[see e.g.][]{Parry08} --- our observations will form additional evidence for, or against, this departure from the expectations of LambdaCDM.

In light of the apparent complexity of the VOD region, additional radial velocities are critically needed. They will enable us to clarify if features $A$, $B$ \& $C$ form a coherent common stream. With revised n-body modelling, it will then be possible to test if these features can be accommodated by an origin in the Sgr debris stream. We aim to obtain further data with the AAT's AAOmega spectrograph. In the near future it will be possible to apply the techniques of the present study to the hemispheric coverage of SkyMapper \citep{Keller07}. 

\section*{Acknowledgments}
This research has been supported in part by the Australian Research Council through Discovery Project Grants DP0343962 and DP0878137.

\end{document}